%% file: main.tex
\documentclass[sigconf]{acmart}
\newif\ifarxiv
\arxivtrue 
\ifarxiv
\renewcommand\footnotetextcopyrightpermission[1]{} % removes footnote with conference information in first column
\fi
\AtBeginDocument{%
  }

\setcopyright{acmcopyright}
\copyrightyear{2023}
\acmYear{2023}
\acmDOI{XXXXXXX.XXXXXXX}

\acmConference[ASP-DAC 2023]{ACM Asia South Pacific Design Automation Conference}{Janurary 16--19,
  2023}{Tokyo, Japan}
\acmPrice{15.00}
\acmISBN{978-1-4503-XXXX-X/18/06}

\usepackage{multirow}
\usepackage{color, colortbl}
\usepackage{amsmath,amsfonts}
\usepackage{algorithmic}
\usepackage{iaik_defaults}
\usepackage{graphicx}
\usepackage{textcomp}
\usepackage{xcolor}
\usepackage{xspace}
\usepackage{changepage}
\usepackage{listings}
\usepackage[most]{tcolorbox}
\usepackage{adjustbox}

\include{results/results}

\newif\ifanonymous
\newcommand\scfi{SCFI\xspace}

\ifanonymous
\newcommand\link{Repository link hidden for the blind review process.}
\else
\newcommand\link{Repository link will be provided after paper acceptance.}
\fi

\definecolor{vgreen}{RGB}{104,180,104}
\definecolor{vblue}{RGB}{49,49,255}
\definecolor{vorange}{RGB}{255,143,102}
\definecolor{lightGray}{gray}{0.9}

\lstdefinestyle{verilog-style}
{
    language=Verilog,
    basicstyle=\small\ttfamily,
    keywordstyle=\color{vblue},
    identifierstyle=\color{black},
    commentstyle=\color{vgreen},
    morekeywords = [2]{
      typedef,
      enum
     }
}
 
\begin{document}
\ifarxiv
\pagestyle{plain}
\fi
\title{SCFI: State Machine Control-Flow Hardening Against Fault Attacks}

\author{Pascal Nasahl}
\email{pascal.nasahl@iaik.tugraz.at}
\affiliation{%
  \institution{Graz University of Technology}
  \city{Graz}
  \country{Austria}
}

\author{Martin Unterguggenberger}
\email{martin.unterguggenberger@lamarr.at}
\affiliation{%
  \institution{Graz University of Technology, Lamarr Security Research}
  \city{Graz}
  \country{Austria}
}

\author{Rishub Nagpal}
\email{rishub.nagpal@lamarr.at}
\affiliation{%
  \institution{Graz University of Technology, Lamarr Security Research}
  \city{Graz}
  \country{Austria}
}

\author{Robert Schilling}
\email{robert.schilling@iaik.tugraz.at}
\affiliation{%
  \institution{Graz University of Technology}
  \city{Graz}
  \country{Austria}
}

\author{David Schrammel}
\email{david.schrammel@iaik.tugraz.at}
\affiliation{%
  \institution{Graz University of Technology}
  \city{Graz}
  \country{Austria}
}

\author{Stefan Mangard}
\email{stefan.mangard@iaik.tugraz.at}
\affiliation{%
  \institution{Graz University of Technology, Lamarr Security Research}
  \city{Graz}
  \country{Austria}
}

\renewcommand{\shortauthors}{Nasahl et al.}

\begin{abstract}
Fault injection~(FI) is a powerful attack methodology allowing an adversary to entirely break the security of a target device.
As finite-state machines~(FSMs) are fundamental hardware building blocks responsible for controlling systems, inducing faults into these controllers enables an adversary to hijack the execution of the integrated circuit.
A common defense strategy mitigating these attacks is to manually instantiate FSMs multiple times and detect faults using a majority voting logic.
However, as each additional FSM instance only provides security against one additional induced fault, this approach scales poorly in a multi-fault attack scenario.

In this paper, we present \scfi: a strong, probabilistic FSM protection mechanism ensuring that control-flow deviations from the intended control-flow are detected even in the presence of multiple faults.
At its core, \scfi consists of a hardened next-state function absorbing the execution history as well as the FSM's control signals to derive the next state.
When either the absorbed inputs, the state registers, or the function itself are affected by faults, \scfi triggers an error with no detection latency.
We integrate \scfi into a synthesis tool capable of automatically hardening arbitrary unprotected FSMs without user interaction and open-source the tool.
Our evaluation shows that \scfi provides strong protection guarantees with a better area-time product than FSMs protected using classical redundancy-based approaches.
Finally, we formally verify the resilience of the protected state machines using a pre-silicon fault analysis tool.

\end{abstract}




\maketitle

%------------------------------------------------------------------------------%
\section{Introduction}
\label{sec:cfi_fsm:introduction}
%------------------------------------------------------------------------------%

Fault attacks are active, physical attacks that allow an adversary to manipulate the execution of a digital circuit.
In these attacks, one or multiple faults are injected into certain gates, wires, or registers of a logical hardware block.
The effects of these faults, which comprise transient bit-flips or permanent stuck-at effects, manipulate the execution of the hardware block and an adversary can exploit this malfunctional behavior~\cite{DBLP:conf/fdtc/VerbauwhedeKS11}.
Finite-state machines~(FSMs) are lucrative fault targets, as these fundamental hardware blocks are responsible of controlling systems and their datapaths.
By hijacking the execution flow of the FSM using faults, an adversary can manipulate the FSM to enter states which cannot be reached from the current state.
Hence, due to the severity of these attacks, security-sensitive state machines need dedicated protection against faults.

A common fault defense strategy is to encode the FSM states such that they are separated with a certain Hamming Distance~\cite{DBLP:conf/wisa/AkdemirHS09, DBLP:conf/date/ChoudhuryFT21, muhtadi2021sparse}.
However, this can only mitigate attackers aiming to induce faults into the state registers.
Other defense strategies~\cite{djordjevic2005concurrent} introduce monitors which check whether the conducted state transition is in the list of valid state transitions.
Leveugle et al.~\cite{DBLP:journals/tc/LeveugleS90} dynamically verifies that the state transitions stay within the intended execution flow, which is determined during synthesis using the control-flow graph~(CFG) of the FSM.
There, on each state transition, a signature is derived, and a monitor checks whether the signature matches the predetermined signature of the CFG.
However, faults induced either into the next-state logic or into the FSM's control signals still enable adversaries to redirect the control-flow within the bounds of the CFG.
Moreover, the fault detection latency of monitor-based schemes is high and the error coverage is often insufficient~\cite{rochet1995efficiency}.

Redundantly instantiating the next-state logic and comparing the resulting states typically requires manual effort by the RTL designer.
Moreover, this approach requires an additional redundant next-state logic for each additional fault protection layer.
Hence, the area overhead of redundancy-based protection mechanisms scales poorly, especially when considering multi-fault attacks, \eg quadruple laser fault injection~\cite{alpha}. 

%------------------------------------------------------------------------------%
\subsection*{Contribution}
%------------------------------------------------------------------------------%

In this paper, we introduce \scfi, a scalable mitigation approach probabilistically protecting the control-flow of finite-state machines against multi-fault attacks.
\scfi ensures that any control-flow deviation from the intended control-flow is detected with a high probability by substituting the unprotected next-state logic of the controller with a fault-hardened next-state logic.
Internally, this hardened logic absorbs the control signals and the execution history and only generates a valid next state when these inputs are not tampered by faults.
When either the control signals, the current state (\ie the execution history), or the next-state logic itself is targeted by faults, the logic ensures that these faults corrupt the next state output to a degree which can be detected.
To ensure this behavior, \scfi uses a lightweight diffusion layer, which is based on a maximum distance separable~(MDS) matrix multiplication.
We integrate \scfi into the Yosys synthesis suite to automatically protect arbitrary FSMs against fault attacks without any user interaction and open-source\footnote{\link} the modified toolchain.
In order to evaluate the area and timing overhead, we synthesized several FSMs used in an industry-driven open-source project with our modified synthesis suite.
Our comparison with a redundancy-based protection approach of the FSM's next-state logic shows that \scfi scales better in terms of area-time product for different fault protection levels than classical redundancy-based protection approaches.
Finally, we utilize a pre-silicon fault analysis tool to formally verify the fault resiliency of the hardened FSMs.

%------------------------------------------------------------------------------%
%\subsection*{Outline}
%------------------------------------------------------------------------------%
%Section~\ref{sec:cfi_fsm:background} gives an introduction to fault attacks and finite-state machines.
%In Section~\ref{sec:cfi_fsm:threat}, we discuss the assumed threat model, provide an attacker description, and formulate the goal of protecting the FSM's next-state function.
%Section~\ref{sec:cfi_fsm:nslogic} gives an overview of the \scfi design and Section~\ref{sec:cfi_fsm:implementation} reveals implementation details of \scfi.
%In Section~\ref{sec:cfi_fsm:eval}, we discuss security properties of \scfi and in Section~\ref{sec:cfi_fsm:ot} we analyze FSMs protected with \scfi.
%Finally, Section~\ref{sec:cfi_fsm:conclusion} concludes this work.

%------------------------------------------------------------------------------%
\section{Background}
\label{sec:cfi_fsm:background}
%------------------------------------------------------------------------------%

This section provides fundamental background on fault attacks and finite-state machines required for the subsequent chapters.

%------------------------------------------------------------------------------%
\subsection{Fault Attacks}
%------------------------------------------------------------------------------%
Fault attacks are commonly used to break the security of embedded devices.
In these physical attacks, one or multiple faults are induced into the circuit, causing several side effects at the electrical level.
These electrical effects comprise timing violations and other disturbances~\cite{DBLP:journals/iacr/Richter-Brockmann21} and they influence the execution of the target.
By exploiting the effects of a fault, an adversary is capable of hijacking the control-flow of software~\cite{DBLP:conf/fdtc/TimmersSW16, DBLP:conf/fdtc/TimmersM17, nasahl2019attacking}, bypassing security measures, such as secure-boot~\cite{DBLP:journals/tc/VasselleTMME20, DBLP:conf/woot/CuiH17}, or extracting secret keys used by cryptographic primitives~\cite{DBLP:conf/crypto/BihamS97, DBLP:journals/tches/DobraunigEKMMP18}.

Originally, fault attacks were pure physical attacks requiring an adversary to have physical access to the target device.
To induce a fault, attackers interrupt the supply voltage or the clock signal, decapsulate the chip and shoot with a laser directly into the die, or use electromagnetic pulses~\cite{DBLP:journals/tvlsi/KaraklajicSV13}.
However, recent publications, such as Plundervolt~\cite{DBLP:conf/sp/MurdockOGBGP20}, CLKSCREW~\cite{DBLP:conf/uss/TangSS17}, or VoltJockey~\cite{DBLP:conf/ccs/QiuWLQ19}, demonstrated that faults also could be induced remotely in software, increasing the attack surface of fault attacks even more.

In general, a fault $f \in F$ is described using the set $K=\{e, s, t\}$ where $e$ is the effect of a fault,
$s$ the spatial, and $t$ the temporal dimension of the fault.
Typically, the fault effect $e$ comprises transient, \ie bit-flips, or stuck-at effects.
The spatial $s$ and temporal $t$ dimensions of a fault describe where (which gate or wire) and when (which clock cycle) a fault is induced.
The set $F$ consists of all possible fault combinations and an adversary typically can inject up to a certain number of faults into the circuit.

%------------------------------------------------------------------------------%
\subsection{Finite-State Machines}
\label{sec:cfi_fsm:fsm}
%------------------------------------------------------------------------------%

Finite-state machines~(FSMs) are sequential circuits responsible for controlling systems and their datapaths.
Internally, an FSM maintains a finite set of states, and a state-transition into the next state that is controlled by the input signals, \ie the control signals and the current state.
The outputs of a Mealy-type FSM are defined by the current state and the input signals, and the outputs of a Moore-type FSM only depend on the current state.

\begin{figure}[h]
  \centering
  \includegraphics[width=0.6\linewidth]{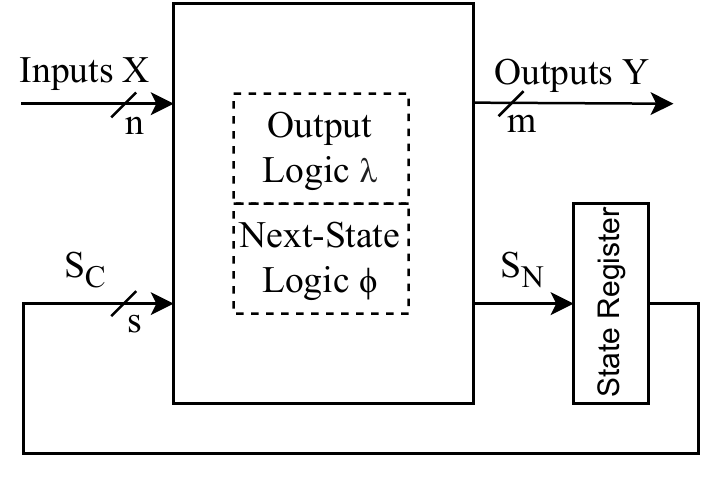}
  \caption{General structure of a state machine.}
  \label{fig:scfi:fsm}
\end{figure}

As depicted in Figure~\ref{fig:scfi:fsm}, an FSM is described using the 5-tuple $\{S, X, Y, \phi, \lambda\}$.
The $|S|$ states of an FSM are represented as a $s$-bit symbol $S$, where the size $s$ needs to be at least $s=\lceil log_2(|S|) \rceil$ bits to comprise the entire state space.
Furthermore, the FSM consists of $n$-bit control signals $X$ and $m$-bit output signals $Y$.
The FSM uses the next-state function $S_N=\phi(X,S_C)$ to derive the next state $S_N$ from the current state $S_C$ and the control signals $X$.
For a Mealy machine, the output $Y$ depends on the current state $S_C$ and the input signals $X$ and is described using the output function $Y=\lambda(X,S_C)$.
\begin{figure}[h]
  \centering
  \includegraphics[width=0.40\linewidth]{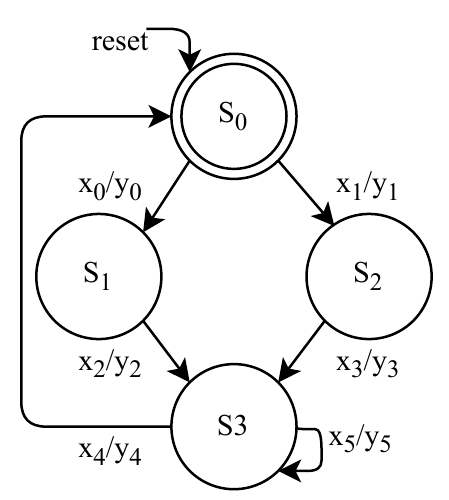}
  \caption{Control-flow graph of an FSM.}
  \label{fig:scfi:cfg}
\end{figure}
The execution-flow of an FSM can be described using a directed graph, as shown in Figure~\ref{fig:scfi:cfg}.
This graph, which is also called a control-flow graph~(CFG), comprises all valid transitions $t \in CFG$ the FSM can perform.
A valid transition is defined by the valid $\{S_C, X\}$ pairs and the given next-state function $\phi$.

%------------------------------------------------------------------------------%
\section{Threat Model}
\label{sec:cfi_fsm:threat}
%------------------------------------------------------------------------------%
We consider a powerful adversary capable of injecting $N-1$ faults in different clock cycles and at different locations into the device under attack.
These faults can be induced independently of the used fault methodology, \ie we consider local and remote injecting techniques.
Similar to related work, we model the impact of a fault as a transient, \ie a bit-flip, or a permanent, \ie a stuck-at, effect.
The spatial dimension of the induced fault comprises wires as well as combinational and sequential elements of the logic.

%------------------------------------------------------------------------------%
\subsection{Attacker Description}
\label{sec:cfi_fsm:attacker}
%------------------------------------------------------------------------------%
Within this threat model, an attacker aims to hijack the execution-flow of a security-sensitive state machine in the circuit.
Based on the general description of a state machine provided in Section~\ref{sec:cfi_fsm:fsm}, the adversary can achieve this goal by inducing faults into the next-state logic.
A fault into the next-state logic allows an adversary to hijack the execution flow of the FSM and to indirectly change the output signals.
This fault target can be modeled using the modified next-state logic $S_N=\phi(S_C, X, F_N)$, where $F_N$ describes one or multiple faults.
Based on this formula, an adversary can induce faults into different fault targets~(FT):\newline
\textit{FT1 State Registers:} A fault into the state registers allows the
\begin{adjustwidth}{0.1cm}{}
 adversary to arbitrarily redirect the control-flow of the FSM inside $t \in CFG$ or outside $t \notin CFG$ the control-flow graph.
For the CFG in Figure~\ref{fig:scfi:cfg}, the adversary could flip bits in the state registers to directly jump from $S_0$ to $S_3$.
\end{adjustwidth}
\textit{FT2 Control Signals:} By inducing bit-flips into the control signals,
\begin{adjustwidth}{0.1cm}{}
 the adversary can manipulate the control-flow of the FSM within the borders of the CFG.
For example, a fault into the control signal $x_0$ or into the comparison logic can hijack the execution $S_0 \rightarrow S_1$ to $S_0 \rightarrow S_2$ in Figure~\ref{fig:scfi:cfg}.
\end{adjustwidth}
\textit{FT3 Next-State Logic:} When directly targeting the logic of the next-
\begin{adjustwidth}{0.1cm}{}
 state function, the adversary can arbitrarily redirect the control-flow of the FSM within or outside the bounds of the CFG.
\end{adjustwidth}

%\end{adjustwidth}
%\textbf{FT2 Output Logic:} A fault into the output-logic allows an
%\begin{adjustwidth}{0.1cm}{}
% adversary to directly change the output signals.
% This fault target can be modeled using the modified output logic $Y=\lambda(S_C, X, F_N)$.
% A fault into the output logic can be induced into the state registers \textit{FT2.1}, the control signals \textit{FT2.2}, the logic \textit{FT2.3}, or into the output signals \textit{FT2.4}.
%\end{adjustwidth}
%------------------------------------------------------------------------------%
\subsection{Goal - Fault Secure FSM}
\label{sec:cfi_fsm:aim}
%------------------------------------------------------------------------------%

In order to comprehensively protect the control-flow of finite-state machines against fault attacks, dedicated fault countermeasures must consider all fault targets \textit{FT1}, \textit{FT2}, and \textit{FT3}.
The goal is that a fault-protected controller $FSM_F$ influenced by faults detects any control-flow deviations from the control-flow of an identical copy $FSM_{\bar{F}}$ which is not affected by faults, \ie $\phi_F(S, X, F_N)$ $=?$ $\phi_{\bar{F}}(S, X, 0)$.

%------------------------------------------------------------------------------%
\section{Design}
\label{sec:cfi_fsm:nslogic}
%------------------------------------------------------------------------------%

To comprehensively protect finite-state machines against control-flow hijacks, with \scfi, we maintain the integrity of the control-flow by introducing a fault-hardened next-state logic $\phi_{FH}$.
This hardened next-state logic prevents that a fault into \textit{FT1}, \textit{FT2}, or \textit{FT3} enables the adversary to redirect the control-flow inside or outside the boundaries of the CFG.
This function $\phi_{FH}$ is internally constructed using a multi-input signature register~(MISR) and it links the entire execution history in a compressed format to detect control-flow deviations.
To enter the next valid state, the execution history as well as the control signals need to be genuine.
\begin{figure}[t]
  \centering
  \includegraphics[width=0.55\linewidth]{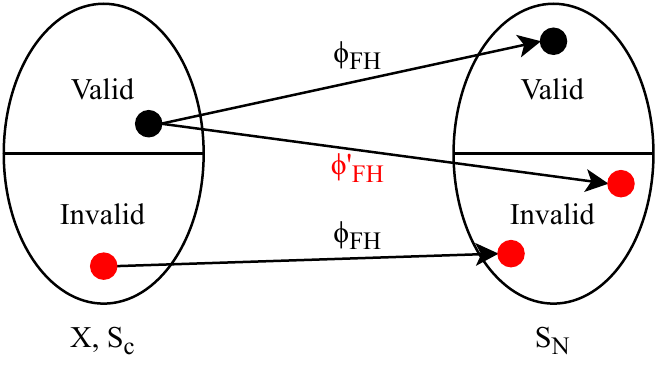}
  \caption{Mapping of valid and invalid input tuples to a valid or invalid next state.}
  \label{fig:scfi:mapping}
\end{figure}
As shown in Figure~\ref{fig:scfi:mapping}, $\phi_{FH}$ maps a valid tuple $\{X,S_C\}$, which includes the execution history in $S_C$, into a valid next state $S_N$.
When an adversary induces faults into \textit{FT1}...\textit{FT3}, \ie either into the tuple $\{X,S_C\}$ or into the $\phi_{FH}$ logic, $\phi_{FH}$ forces the FSM into a non-escapable terminal error state.
\begin{figure}[h]
\centering
\begin{tcolorbox}[sidebyside,
sidebyside align=top,
standard jigsaw,
opacityback=0,
arc=0pt,outer arc=0pt,
boxrule=1pt,
left = 0pt,
top = 0pt,
bottom = 0pt,
%blankest % uncomment to remove colorbox
]
\begin{lstlisting}[style={verilog-style},basicstyle=\footnotesize\linespread{0.5}]
unique case(SC)
 S0: begin
  SN = S0;
  if(x0)
   SN = S1;
  else if(x1)
   SN = S2;
 end
 S1: begin
  SN = S1;
  if(x2)
   SN = S3;
 end
\end{lstlisting}
\tcblower
\begin{lstlisting}[style={verilog-style},basicstyle=\footnotesize\linespread{0.5},mathescape=true]
unique case(SC)
 S0: begin
  SN = $\phi_{FH}(SC, X)$;
 end
 S1: begin
  SN = $\phi_{FH}(SC, X)$;
 end
 ERROR: begin
  SN = ERROR;
 end
 default: begin
  fsm_alert = err_signal;
  SN = ERROR;
 end
\end{lstlisting}
\end{tcolorbox}
\caption{Unprotected and protected next-state logic of an example FSM.}
\label{fig:scfi:fsmprotunprot}
\end{figure}
Figure~\ref{fig:scfi:fsmprotunprot} depicts the transformation of an unprotected next-state logic of an example FSM into a protected version.
The unprotected FSM is susceptible to faults, as a single fault into the state registers, the comparison logic, or the control signals can change the execution-flow of the FSM.
\scfi closes these attack vectors by deriving the next state using $\phi_{FH}$.
If the current state, the control signals, or the next-state logic is tampered with a fault, $\phi_{FH}$ produces an invalid state and enters the non-escapable default error state.
To achieve this protection degree, the next-state function and its inputs and outputs need to fulfill requirements \textbf{R1} to \textbf{R3}:\newline
\textbf{R1 Encoded Control Signals:} All control signals $X$ are encoded
\begin{adjustwidth}{0.1cm}{}
 to $X_e$. The encoding needs to guarantee that the attacker needs at least $N$ bit-flips to manipulate a valid control-signal codeword to another valid codeword.
\end{adjustwidth}
\textbf{R2 Encoded States:} All states $S$ are encoded to $S_e$. Similar
\begin{adjustwidth}{0.1cm}{}
 to the control signals, the encoding needs to guarantee a minimum Hamming Distance between valid states of $N$.
\end{adjustwidth} 
\textbf{R3 Hardened Next-State Function:} The hardened next-state
\begin{adjustwidth}{0.1cm}{}
 function $\phi_{FH}$ generates an encoded next state $S_{Ne}$ fulfilling \textbf{R2} for each encoded control signal and encoded current state tuple $\{S_{Ce}, X_e\}$. 
 Moreover, $\phi_{FH}$ needs to ensure that up to $N-1$ bit-flips into its circuit or into the input space affect the output in such a way, that the faults can be detected, \ie an invalid state $S_{Ne}$ is generated.
\end{adjustwidth} 
Due to requirement \textbf{R3}, the state derived in different paths merging at some point also produces different encoded states.
For example, the path $S_1 \rightarrow S_3$ in Figure~\ref{fig:scfi:cfg} derives a different state than the path $S_2 \rightarrow S_3$. 
As maintaining different state symbols for a single state is costly, we add an additional requirement:\newline
\textbf{R4 Collision Capability:} The hardened next-state function
\begin{adjustwidth}{0.1cm}{}
  needs to produce the same encoded next state for different paths using a modifier, \ie $\phi_{FH}(S_{C1e}, X_{1e}, Mod_1)$ == $\phi_{FH}(S_{C2e}, X_{2e}, Mod_2)$ for $S_{C1e} \neq S_{C2e}$ and $X_{1e} \neq X_{2e}$. 
 The modifiers $Mod_1$ and $Mod_2$ are used to produce a state collision.
\end{adjustwidth}
%------------------------------------------------------------------------------%
\subsection{Selection of the Hardened Next-State Function}
%------------------------------------------------------------------------------%

\begin{figure}
  \centering
  \includegraphics[width=0.8\linewidth]{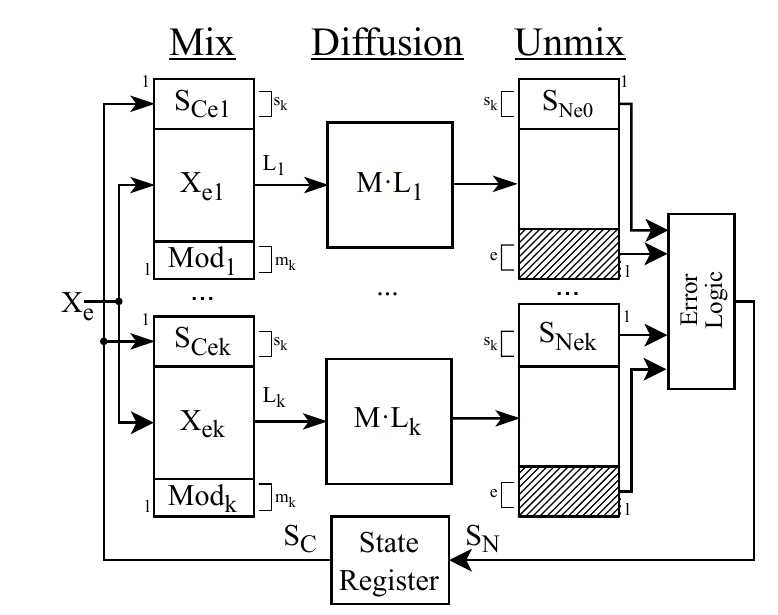}
  \caption{SCFI hardened next-state function.}
  \label{fig:scfi:mdsblock}
\end{figure}

In \scfi, we utilize a lightweight diffusion function used in cryptographic primitives for the hardened next-state function $\phi_{FH}$. 
The properties of this function imply that any fault at the input or within the logic maximally affect the output, thereby substantially decreasing probability of a successful fault attack.
Figure~\ref{fig:scfi:mdsblock} depicts the structure of $\phi_{FH}$ mapping the input space $\{S_{Ce}, X_e, Mod\}$ to a next encoded state $S_{Ne}$.

\paragraph{Mix Layer.}
In this layer, the input triple is split into $k$ $l$-bit vectors $L$.
For this, the encoded current state, the encoded control signals, and the modifier are split into $k$ shares and each share is placed into the vectors, as shown in Figure~\ref{fig:scfi:mdsblock}.

\paragraph{Diffusion Layer.}
Then, in the diffusion layer, the vectors $L$ are absorbed by $k$ diffusion functions.
These functions conduct a linear transformation $D(L) = M\cdot L$ which is a matrix multiplication of vector $L$ with matrix $M$ in a specific field.
This transformation, depending on the choice of matrix $M$, yields a strong diffusion.
Ideal choices of this matrix are called maximum distance separable~(MDS) matrices, maximizing the diffusion property.

\paragraph{Unmix Layer.}
The output of the diffusion layer is stored into $k$ $l$-bit vectors.
The concatenation of the first $k$-bits of each output vector results in the encoded next state $S_{Ne}$.
As the size $k\cdot l$ of the output space is larger than the size $s_e$ of the encoded state, $k\cdot l - s_e$ bits are free.
\scfi uses, depending on the required fault security, the $e$ topmost bits of each output vector as error detection bits $E$.
Here, by choosing a corresponding modifier $Mod$, $\phi_{FH}$ sets these bits to a predefined value, \ie $1$.
In the error logic, the logical \texttt{AND} of $S_{Ne}$ and $E$ infects the next state when a fault-induced error happens.

%------------------------------------------------------------------------------%
\section{Implementation}
\label{sec:cfi_fsm:implementation}
%------------------------------------------------------------------------------%

We open-source a modified version of the Yosys~\cite{Yosys} open synthesis suite capable of automatically protecting arbitrary FSMs with \scfi.
The protection can be enabled globally or selectively for the unprotected FSMs in the design flow with a certain fault protection level $N$.
Our implementation adds a new Yosys pass to the suite operating in between of other optimization passes before the design is mapped to the logic gate level.
Note that the RTL designer only needs to manually encode the control signals with a Hamming Distance of $N$-bits in the modules driving these signals.

%------------------------------------------------------------------------------%
\subsection{Next-State Logic}
%------------------------------------------------------------------------------%

First, our custom FSM protection pass identifies the unprotected FSM by utilizing the existing Yosys FSM passes.
Then, the FSM's state variables are re-encoded so that the Hamming Distance between these variables is $N$.
Afterwards, our pass extracts the CFG of the FSM and stores the current state, the next state, and the control signals for each control-flow edge.
With this information, the modifier $Mod$ for state transition is determined, satisfying the equation $MDS(S_{Ce},X_e, Mod)=S_{Ne}$.

For the MDS diffusion function, we use a lightweight construction with a minimal gate count proposed by Duval et al.~\cite{DBLP:journals/tosc/DuvalL18}.
\begin{figure}[t]
  \centering
  \includegraphics[width=0.37\linewidth]{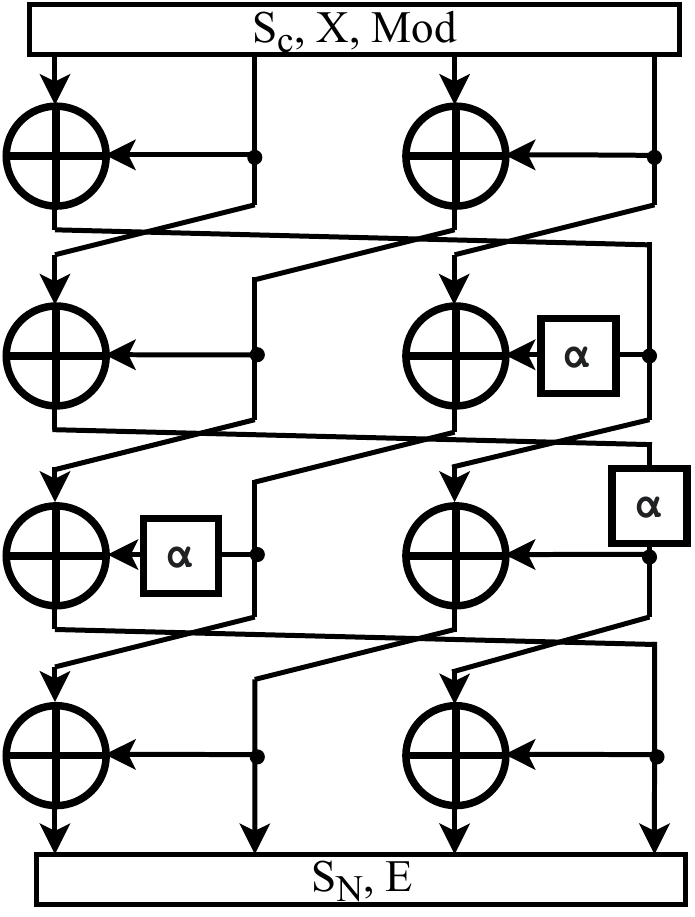}
  \caption{Internal structure of the MDS matrix multiplication~\cite{DBLP:journals/tosc/DuvalL18}. All elements operate on 1-bytes each.}
  \label{fig:scfi:mdsmatrixmult}
\end{figure}
As shown in Figure~\ref{fig:scfi:mdsmatrixmult}, this function splits the 32-bit input space into 4 8-bit chunks, performs the matrix multiplication, and returns 4 8-bit vectors which form the 32-bit output.
In \scfi, we selected the $M\begin{smallmatrix}8, & 3\\ 4, & 6 \end{smallmatrix}$~\cite{DBLP:journals/tosc/DuvalL18} MDS matrix operating in the field $\mathbb{F}_2[\alpha]$ with $\alpha=X^8 + X^2 + 1$. 
This particular matrix has a low \texttt{XOR} count with a slightly larger logical depth compared to other matrices in the $4\times 4$ category. 
We note that the choice of MDS matrix can be changed according to design requirements, \ie area or timing constraints.

\begin{figure}[h]
  \centering
  \includegraphics[width=0.95\linewidth]{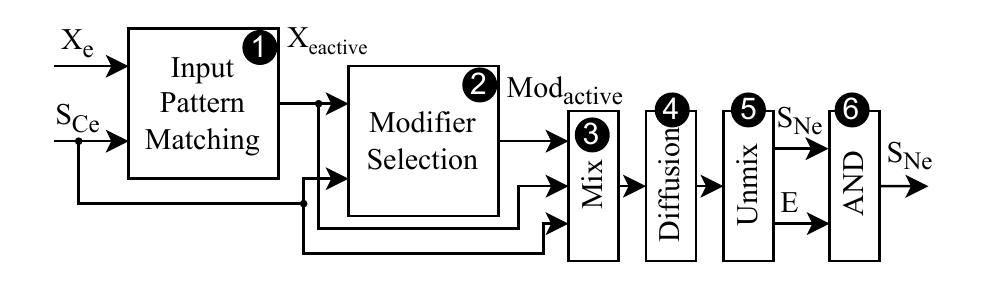}
  \caption{The next-state logic hardening pass.}
  \label{fig:scfi:nsimplementation}
\end{figure}

Having the modifiers, our pass describes the logic of the next-state function in the internal Yosys register-transfer level intermediate language~(RTLIL).
As depicted in Figure~\ref{fig:scfi:nsimplementation}, first~\circledb{1}, the active control signal $X_{e_{active}}$ is determined by performing a pattern match of the control signal and the current state $S_{Ce}$.
Then, using this signal and $S_{Ce}$, the modifier for this input is selected~\circledb{2}.
In the mix layer~\circledb{3}, the wires of the triple $\{S_{Ce}, X_{e_{active}}, Mod_{active}\}$ are distributed to the $k$ 32-bit input MDS diffusion functions.
These lightweight diffusion functions~\circledb{4} consist of only \texttt{XOR} gates.
In the unmix layer~\circledb{5}, the next state $S_{Ne}$ is concatenated and the error bits $E$ are selected.
By connecting $S_{Ne}$ and $E$ using \texttt{AND} gates~\circledb{6}, a fault infectively destroys the next state.

%After the execution of the FSM protection pass, the hardened next-state function described in RTLIL is mapped to logic gates by the subsequent Yosys passes.

%------------------------------------------------------------------------------%
\section{Evaluation}
\label{sec:cfi_fsm:ot}
%------------------------------------------------------------------------------%
To evaluate the effectiveness of \scfi in terms of area, timing, and security when protecting security-sensitive FSMs of an industry-driven project, we integrate our custom Yosys pass into the design flow of the OpenTitan~\cite{johnson2018titan} secure element.
This chip, which is entirely open-source, acts as a secure root-of-trust and provides a key storage and cryptographic accelerators.

\begin{table}[b]
\caption{Area overhead for protecting different FSMs using redundancy or SCFI.}
\label{tab:cfi_fsm:evaluation}
\begin{adjustbox}{max width=0.95\columnwidth}
\begin{tabular}{lccccccc}
\hline
                     & Unprotected    & \multicolumn{3}{c}{Redundancy}    & \multicolumn{3}{c}{SCFI}          \\
                     & Area {[}GE{]} & \multicolumn{3}{c}{Area {[}\%{]}} & \multicolumn{3}{c}{Area {[}\%{]}} \\
\multicolumn{1}{r}{Protection Level} &                 & 2                 & 3                   & 4                  & 2                 & 3                   & 4          \\ \hline \rowcolor{lightGray}
adc\_ctrl\_fsm       & \adcctrlfsmBase     & \adcctrlfsmRedTwo     & \adcctrlfsmRedThree     & \adcctrlfsmRedFour     & \adcctrlfsmCfiTwo     & \adcctrlfsmCfiThree     & \adcctrlfsmCfiFour     \\ 
aes\_control         & \aescontrolBase     & \aescontrolRedTwo     & \aescontrolRedThree     & \aescontrolRedFour     & \aescontrolCfiTwo     & \aescontrolCfiThree     & \aescontrolCfiFour     \\ \rowcolor{lightGray} 
i2c\_fsm             & \icfsmBase          & \icfsmRedTwo          & \icfsmRedThree          & \icfsmRedFour          & \icfsmCfiTwo          & \icfsmCfiThree          & \icfsmCfiFour          \\ 
ibex\_controller     & \ibexcontrollerBase & \ibexcontrollerRedTwo & \ibexcontrollerRedThree & \ibexcontrollerRedFour & \ibexcontrollerCfiTwo & \ibexcontrollerCfiThree & \ibexcontrollerCfiFour \\ \rowcolor{lightGray} 
ibex\_lsu            & \ibexlsuBase        & \ibexlsuRedTwo        & \ibexlsuRedThree        & \ibexlsuRedFour        & \ibexlsuCfiTwo        & \ibexlsuCfiThree        & \ibexlsuCfiFour        \\ 
otbn\_controller     & \otbncontrollerBase & \otbncontrollerRedTwo & \otbncontrollerRedThree & \otbncontrollerRedFour & \otbncontrollerCfiTwo & \otbncontrollerCfiThree & \otbncontrollerCfiFour \\ \rowcolor{lightGray} 
pwrmgr\_fsm          & \pwrmgrfsmBase      & \pwrmgrfsmRedTwo      & \pwrmgrfsmRedThree      & \pwrmgrfsmRedFour      & \pwrmgrfsmCfiTwo      & \pwrmgrfsmCfiThree      & \pwrmgrfsmCfiFour      \\ \hline
Geometric Mean       &                     & 17.5 & 42.9 & 67.6 & 9.6 & 21.8 & 27.1 \\ \hline
\end{tabular}
\end{adjustbox}
\end{table}
%------------------------------------------------------------------------------%
\subsection{Area Overhead}
%------------------------------------------------------------------------------%
In order to evaluate the area overhead introduced by \scfi, we analyzed unprotected~\textit{(i)}, manually protected~\textit{(ii)}, and automatically protected~\textit{(iii)} FSMs.
As the reference~\textit{(i)} for our evaluation, we selected several FSMs of OpenTitan and synthesized the entire corresponding module with Yosys using the open-source Nangate45 standard cell library.
For the manually protected~\textit{(ii)} FSMs, we encoded the control signals with a Hamming Distance of $N$-bits and instantiated the next-state logic of the FSM $N$ times.
To detect control-flow hijacks triggered by faults, we designed a small error logic monitoring the state registers of the redundant FSMs and raising an error signal when one or more state values mismatch.
Finally, we automatically protected~\textit{(iii)} the reference~\textit{(i)} FSMs by calling the \scfi Yosys pass in the design flow.
Similar to the manually protected~\textit{(ii)} FSMs, we encoded the control signals with a HD of $N$-bits and configured \scfi that at least $N$ faults are required to hijack the FSM.
Table~\ref{tab:cfi_fsm:evaluation} illustrates the area overheads for the three configurations for different FSMs and different protection levels $N$ ranging from $2$ to $4$.
For the manual redundancy approach, the geometric mean of the area overhead is $17.5\,\%$ for $N=2$, $42.9\,\%$ for $N=3$, and $67.6\,\%$ for $N=4$.
In comparison, the geometric mean area overhead for the FSMs protected with \scfi is $9.6\,\%$ for $N=2$, $21.8\,\%$ for $N=3$, and $27.1\,\%$ for $N=4$.
Note that for smaller input spaces $\{S_{Ce},X_e, Mod\}$ the area overhead for \scfi could be higher than for a redundancy approach (\cf \texttt{otbn\_controller} in Table~\ref{tab:cfi_fsm:evaluation}) as \scfi needs to instantiate a MDS matrix with a 32-bit input.

%------------------------------------------------------------------------------%
\subsection{Timing Overhead}
%------------------------------------------------------------------------------%
\scfi affects the timing of the next-state logic by introducing the fault-hardened next-state function $\phi_{FH}$.
However, the timing overhead is minimal, as the logical depth of $\phi_{FH}$ comprises four \texttt{XOR} layers for the MDS multiplication and an \texttt{AND} layer for the error masking.
We successfully synthesized all modules in all configurations depicted in Table~\ref{tab:cfi_fsm:evaluation} for OpenTitan's target frequency of $125\,$MHz with Yosys and the open-source standard cell library.

\begin{figure}
  \centering
  \includegraphics[width=\linewidth]{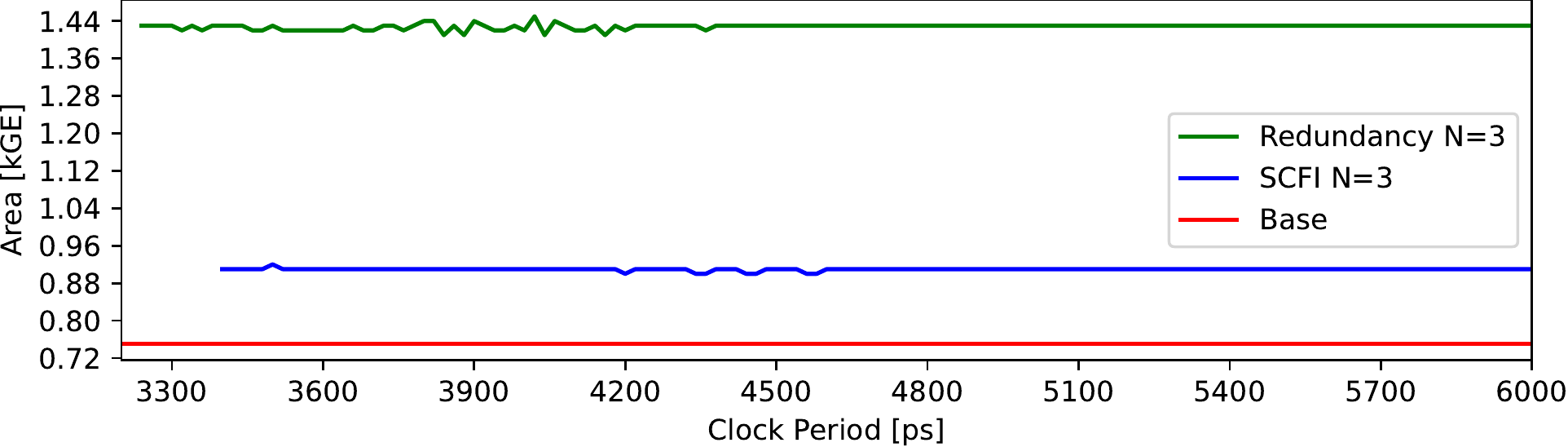}
  \caption{Area-time product for the \texttt{adc\_ctrl\_fsm} module in different configurations.}
  \label{fig:scfi:atproduct}
\end{figure}

Figure~\ref{fig:scfi:atproduct} illustrates the area-time (AT) product for the unmodified, the redundancy-protected, and the \scfi-hardenend \texttt{adc\_ctrl\_fsm} module.
In this plot, we increased the clock period from $3200\,$ps to $6000\,$ps and measured the area in kGE of the design synthesized by Cadence Genus and a proprietary cell library.
For this experiment, we switched from Yosys to the Cadence synthesis suite as Yosys and the internally utilized yosys-abc tool only provides basic area and time optimization functionality.
As shown in Figure~\ref{fig:scfi:atproduct}, Cadence was able to meet the timing for a maximum frequency of $312\,$MHz for the base design, $308\,$MHz for the design using redundancy, and $294\,$MHz for \scfi.
However, this slightly decreased frequency is typically not problematic, as the critical path of a design is usually not in an FSM.
Moreover, as depicted, \scfi achieves a better AT product for protecting the next-state logic of the FSM in the \texttt{adc\_ctrl\_fsm} module than the redundancy approach.

%------------------------------------------------------------------------------%
\subsection{Security Evaluation}
\label{sec:cfi_fsm:eval}
%------------------------------------------------------------------------------%
By encoding the control signals and the state variable, an adversary cannot hijack the state machine by inducing faults into fault targets \textit{FT1} and \textit{FT2}.
As the input pattern matching logic~\circledb{1} of $\phi_{FH}$ operates on these encoded signals, the attacker needs to induce $N$ faults into this block to manipulate the active, encoded control signal.
While a fault into the modifier selection block~\circledb{2}, which consists of multiplexers, could select a different modifier, the attacker cannot exploit this injected fault.
More specifically, a fault would yield a combination of control signal, state, and modifier which creates a non-valid next state.
Internally, the mix layer~\circledb{3} consists of a rewiring of the encoded control signals, the state, as well as the modifier.
Hence, this layer can resist up to $N-1$ faults.
The idea of the diffusion layer~\circledb{4} is that a small change at the input causes a significant change at the output, \ie the avalanche effect.
To achieve this property, \scfi internally uses MDS matrix multiplication yielding optimal diffusion guarantees.
These MDS matrices propagate a bit-flip in a single input byte to all four output bytes, \ie they have a branch number of $5$.
Hence, one or multiple bit-flips into the input triple $\{S_{Ce}, X_{e_{active}, Mod_{active}}\}$ propagate through this function affecting multiple output bits.
By effecting the next state $S_{Ce}$ or the error bits $E$, an invalid state is generated in the unmix~\circledb{5} and error~\circledb{6} layer and the FSM enters the default error state. 
Precisely, there are only $|S_{Ne}| + |E|$ valid output states; an attacker who induces $N$ faults on the next-state function inputs, $\{X, S_{C}\}$, would have a success probability of $P = \frac{|S_{Ne}| + |E|}{k\cdot 2^{32 - (|S_{Ne}| + |E|)}}$.
However, considering that $|S_{Ce}| + |E| << k\cdot 2^{32 - (|S_{Ne} + |E|)}$, the success probability is very small.
For attacks within the next-state function, the MDS property of the diffusion layer ensures that the success probability still remains quite low, albeit it is higher than the previous case.
As shown in Figure~\ref{fig:scfi:mdsmatrixmult} depicting the construction of the MDS matrix, faults in the first three \texttt{XOR} layers propagate to at least two output bytes.
Although a fault at the last layer only affects one output byte, all valid output states $S_N$ are still encoded with a Hamming Distance of $N$, requiring that the adversary needs to induce $N$ bit-flips.

%------------------------------------------------------------------------------%
\subsection{Formal Security Analysis}
\label{sec:cfi_fsm:synfi}
%------------------------------------------------------------------------------%

We formally analyzed the resilience of the diffusion layer consisting of the MDS matrix multiplication by utilizing SYNFI~\cite{DBLP:journals/corr/abs-2205-04775}, a recently introduced pre-silicon fault analysis tool operating at the netlist.
For the analysis, we synthesized an FSM with $14$ state transitions and configured \scfi with a protection level of $2$ bits (HD).
We used SYNFI to analyze whether it is possible to hijack one of the state transitions and enter another next state using faults.
In total, we injected $7644$ single bit-flips exhaustively into all available gates in the MDS matrix multiplication and $32$ ($0.42\,\%$) of these faults enable an adversary to hijack the execution-flow of the FSM.

Note that analyzing the resilience of FSMs against faults is also necessary when using other protection approaches.
For example, when redundantly instantiating the next-state logic to mitigate faults, a synthesis tool aiming to meet timing and area constraints could weaken the security when optimizing the design. 

%------------------------------------------------------------------------------%
\section{Limitation \& Future Work}
\label{sec:cfi_fsm:limitation}
%------------------------------------------------------------------------------%
A potential future work could extend \scfi to adapt the MDS matrix size to the size of the $\{S_C, X, Mod\}$ input triple to further improve the area-time product.
In addition, the formal analysis could be integrated into the Yosys pass to increase security guarantees of \scfi.
Finally, a future work could investigate how \scfi could be extended to also provide protection for the output logic.

A limitation of the current prototype implementation is that the selector signals of the MUXes used in the input pattern matching logic~\circledb{1} are 1-bit signals.
This would allow an adversary to redirect the control-flow within the bounds of the CFG.
To mitigate this attack vector, an updated version of the \scfi Yosys pass could introduce encoded selector signals.

%------------------------------------------------------------------------------%
\section{Conclusion}
\label{sec:cfi_fsm:conclusion}
%------------------------------------------------------------------------------%
In this paper, we presented \scfi, a methodology capable of protecting the control-flow of finite-state machines against fault attacks.
\scfi substitutes the next-state logic of FSMs with a fault-hardened function only deriving a next valid state in a fault-free scenario.
We integrated \scfi into the Yosys synthesis suite and open-sourced our modified toolchain.
Our evaluation shows that the area overhead for FSMs protected with \scfi is lower than for traditional protection approaches.

\ifanonymous
\else
%------------------------------------------------------------------------------%
\section*{Acknowledgments}
%------------------------------------------------------------------------------%
This project has received funding from the Austrian Research Promotion Agency (FFG) via the AWARE project (grant number 41091245).

\fi

\bibliographystyle{ACM-Reference-Format}
\bibliography{bibliography}

\end{document}

%% file: results/results.tex
\newcommand{\adcctrlfsmBase}{1019}
\newcommand{\adcctrlfsmCfiTwo}{14}
\newcommand{\adcctrlfsmCfiThree}{27}
\newcommand{\adcctrlfsmCfiFour}{42}
\newcommand{\adcctrlfsmRedTwo}{38}
\newcommand{\adcctrlfsmRedThree}{76}
\newcommand{\adcctrlfsmRedFour}{121}
\newcommand{\aescontrolBase}{632}
\newcommand{\aescontrolCfiTwo}{6}
\newcommand{\aescontrolCfiThree}{22}
\newcommand{\aescontrolCfiFour}{32}
\newcommand{\aescontrolRedTwo}{13}
\newcommand{\aescontrolRedThree}{44}
\newcommand{\aescontrolRedFour}{77}
\newcommand{\icfsmBase}{2729}
\newcommand{\icfsmCfiTwo}{20}
\newcommand{\icfsmCfiThree}{21}
\newcommand{\icfsmCfiFour}{27}
\newcommand{\icfsmRedTwo}{38}
\newcommand{\icfsmRedThree}{70}
\newcommand{\icfsmRedFour}{109}
\newcommand{\ibexcontrollerBase}{537}
\newcommand{\ibexcontrollerCfiTwo}{13}
\newcommand{\ibexcontrollerCfiThree}{34}
\newcommand{\ibexcontrollerCfiFour}{43}
\newcommand{\ibexcontrollerRedTwo}{29}
\newcommand{\ibexcontrollerRedThree}{75}
\newcommand{\ibexcontrollerRedFour}{122}
\newcommand{\ibexlsuBase}{933}
\newcommand{\ibexlsuCfiTwo}{2}
\newcommand{\ibexlsuCfiThree}{13}
\newcommand{\ibexlsuCfiFour}{16}
\newcommand{\ibexlsuRedTwo}{10}
\newcommand{\ibexlsuRedThree}{21}
\newcommand{\ibexlsuRedFour}{32}
\newcommand{\otbncontrollerBase}{2857}
\newcommand{\otbncontrollerCfiTwo}{5}
\newcommand{\otbncontrollerCfiThree}{5}
\newcommand{\otbncontrollerCfiFour}{6}
\newcommand{\otbncontrollerRedTwo}{1}
\newcommand{\otbncontrollerRedThree}{4}
\newcommand{\otbncontrollerRedFour}{5}
\newcommand{\pwrmgrfsmBase}{301}
\newcommand{\pwrmgrfsmCfiTwo}{33}
\newcommand{\pwrmgrfsmCfiThree}{71}
\newcommand{\pwrmgrfsmCfiFour}{84}
\newcommand{\pwrmgrfsmRedTwo}{89}
\newcommand{\pwrmgrfsmRedThree}{184}
\newcommand{\pwrmgrfsmRedFour}{334}